# Acoustic 'distributed source' mixing of smallest fluid volumes

A. Rathgeber, M. Wassermeier
*Advalytix AG, Brunnthal, Germany*
and
A. Wixforth
*Experimentalphysik I, University of Augsburg, Augsburg Germany*

Acoustically driven mixing of small fluid volumes is reported. Using surface acoustic waves on a mixer chip, conversion of those into bulk waves, and employing wave guiding effects enables us to distribute a set of "virtual sources" for acoustic streaming over large areas. we demonstrate the applicability of our mixing technique to micro array applications, for mixing of individual wells in a micro titer plate, and other state-of-the-art hybridization assays.

Mixing of smallest amounts of fluids is usually a delicate task. For liquids being confined to small volumes, the low Reynold's number [1] is equivalent to an increase of the apparent viscosity. As for mixing applications turbulent streaming which only occurs at high Reynold's numbers (>2000) is advantageous [1], this in turn leads to insufficient mixing in microfluidic systems, as only laminar flow processes are allowed.

Hence, the only way for small fluid volumes to effectively mix is driven by diffusion. Here, the smallness of the system is in favour of the diffusion limited time scales, as the respective length scales are equally small. However, for many applications, especially micro array based assays [2], [3], a deliberate and controlled agitation of the fluid under investigation would be of great importance. This is due to the fact that diffusion driven mixing is effective only on a short range. Micro array hybridization, however, demands for transport molecules in the fluid over large distances. Here, agitation is necessary.

In micro array applications, usually hundreds or thousands of ‚spots' containing a specific DNA sequence or more generally a specific oligonucleotide are deposited in a matrix like manner, usually on a microscope slide or similar carrier.

Sample fluid containing for instance unknown sample oligonucliotides of unknown concentration is then spread across this micro array and brought to hybridization conditions. This usually means elevated temperatures and controlled athmospheric conditions to prevent the evaporation of the precious sample. To save sample volume or to avoid unnecessary high dilution, usually very small sample volumes are created by spreading the sample across the micro array employing a narrow capillary slot. This capillary slot is then defined by a cover glass located on top of the micro array at a defined distance provided by an appropriate spacer material.
In Fig. 1, we depict such a typical arrangement consistsing of a microarray (a), the cover slide (b), and the sample fluid (c) in between.

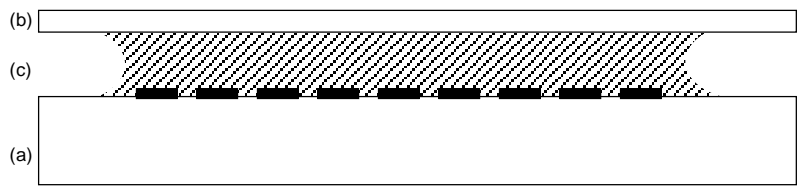

**Fig. 1** : *Schematic of a microarray hybridization setup as described in the text. An array of functionalized spots (indicated by the small black rectangles) containing target molecules is deposited on a slide (a). A thin fluid layer (c) containing the sample molecules is spread over the spotted area of the microarray. A cover slide (b) is used to define the height of the fluid layer and to confine the sample volume.*

The substrate, containing a large number of spots defines the container for the target mollecules. A thin (approx. 50..100 microns) layer of buffer solution containing the probe molecules is covering the spotted area, and a thin cover slide on top of the liquid film defines the closed volume of the assay under consideration.
A quite straight forward way of sample agitation is followed by external pumping. Here, thin tubes or channels are connected to the substrate slide, for example, and either continous or pulse mode pumping to the liquid is used to agitate the fluid and hence the sample molecules to increase the speed of hybridization. Other attempts rely on a (periodic) movement of the substrate slide with respect to the cover lid, hence aiming towards a periodic movement of the taret spots with respect to the sample solution. However, both attemps, as well as similar approaches still do not overcome the ‚problem' of the laminar flow in such thin liquid films. Moreover, external pumping induces an unwanted ‚dead volume' to the process, as some liquid containing precious sample is residing in the external pumping unit or tubes without participating to the hybridization process.

Recently, we have applied the idea of acoustic streaming [4] to the microarray hybridization process [5], [6]. Instead of using external pumps or complicated schemes to move the spotted slide against the liquid and the cover lid, the ‚pumps' are part of the experiment, being in direct contact to the sample fluid under consideration. Here, we use surface acoustic waves (SAW), being generated on a mixer chip, to effectively radiate acoustic energy into the sample. SAW are excited using so-called interdigital transducers [7] on a piezoelectric substrate. In Fig. 2, we schematically show this energy transfer process from a SAW to a liquid layer on top of the substrate.

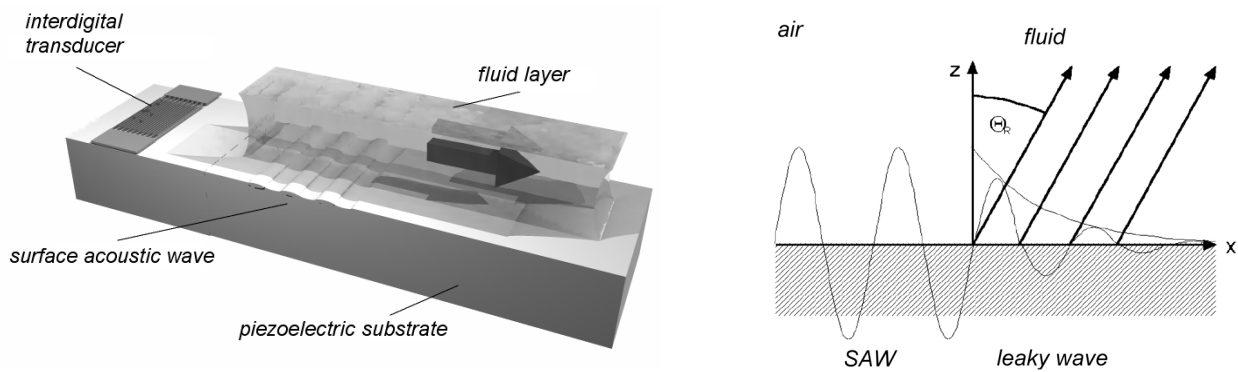

**Fig. 2 :** *Schematic representation of the energy transfer between a SAW and a fluid layer on top of the substrate. A Surface acoustic wave is excited using an interdigital transducer (left) and radiates energy into the fluid. Phase matching conditions require the radiation under an angle _ with respect to the surface of the substrate.*

In our case, the substrate is a LiNbO3 chip, being glued into a glass slide acting as the cover lid in a hybridization experiment. Special fluidic arrangement like small filling holes and surface functionalization of this ‚mixer card' [5],[6] allows for a defined and deliberate filling of the gap between microarray and cover slide without inducing any dead volume. The height of the gap and hence the volume used in the experiment is defind by, e.g., flat spacers being attached to the mixer card. The SAW is excited on the mixer card by application of a high frequency signal (approx. 150 MHz) to the transducers on the chip. Different transducers operating at slightly different frequencies can be addressed electronically to excite different SAW at different locations of the chip, and hence creating different sources for the acoustic streaming in the fluid above.

It can be shown that the absorption of acoustic energy from the SAW to the liquid is very efficient by coupling to longitudinal pressure waves in the liquid. As the sound velocity in the fluid is usually much different from the one in the solid carrying the SAW, phase matching conditions require radiation of the Sound wave in the fluid under an angle with respect to the plane of the mixer card and the micro array substrate respectively. This angle is given by a refraction law, and hence by the ratio of the sound velocities in the fluid and the substrate, respectively. However, as the SAW power is strongly absorbed by the fluid, the sources for the acoustically induced streaming within the fluid are localized either to the source of the SAW itself (the IDT, for example), or to the three phase boundary, where the SAW on a solid substrate enters the fluid (see Fig. 2, right). The arising total flow pattern is hence governed both by a local actuation and viscous damping in the fluid layer.

However, continuity of the mass flow also reqires closed material folding lines in the liquid, which in turn leads to a more or less homogenous agitation of the liquid on top of the microarray. This agitation is more efficient if the fluidic layer is thicker, as its interaction between the microarray substrate and the cover lid (here the mixer card) becomes less pronounced in this case. To account for the exponential decay of the fluid flow as a function of the distance from the exciting transducer, a typical mixer card holds different sound sources (transducers) at different locations on the chip, and larger mixer cards hold more than one single mixer chip.
Hence, there are multiple sources for ultrasound distributed over the spotted area of the microarray.

In this manuscript, we wish to report on a different approach for acoustically induced streaming in a thin layer of liquid spread over a microarray by distributing a manifold of acoustic sources across the spotted area. Aprt from ‚simple' microarrays the very same approach can be evenly successful be applied to different schemes for hybridization, like micro titer plates (MTP), or various other sample holders for such biological assays. For example, instead of employing conventionally spotted micro arrays, recently, multiwell slides as an in-between approach are used by defining small volumes of specific fluids by employing a punched spacer foil between the substrate and the cover slide. Here, the cavities containing the sample liquid are simply defined by the openings in the spacer. Although here, too, in principle our SAW induced streaming can be sucessfully applied (for instance by attaching one or more sound transducers to the specific location of a micro well in a MTP, or at the locations of the small cavities in a spacer defined sample holder, we were able to show that a distribution of sources of ultrasonic energy can be also achieved making use of internal reflections of sound in a cover slide.

The basic idea behind this approach is to excite ultrasound at some location of the mixer card, and to radiate acoustic energy into it being then reflected at the boundaries of the card. If the sound velocity in the mixer card is different from the one in the generating chip, a bulk wave is excited in the card again under an angle with respect to the plane of the piezoelectric generator. In Fig. 3, we depict this idea:

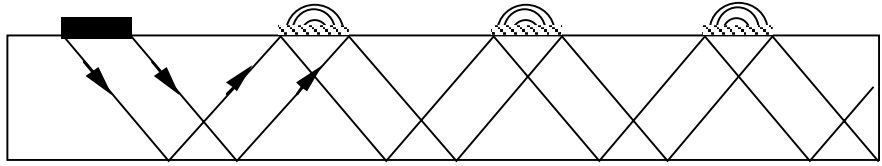

**Fig. 3 :** *Basic idea behind the creation of distributed `virtual sources' to generate acoustic streaming in a fluid in vicinity of the substrate or the mixer card. A sound wave is generated at the transducer indicated to the left in black. Ultrasonic energy in form of a sound wave is radiated into the bulk of the ‚mixer card' under an oblique angle. Imdedance discontinuities at the boundaries of the mixer card cause internal reflections. The points of reflection then act as sources for the emission of ultrasonic energy(schematically indicated by the concentric circles) into a fluid layer.*

The points of reflection of the ultrasound at the boundaries of the mixer card act as efficient sources of radiation of acoustic energy in to a fluid at these boundaries by allowing for a leaking of ultrasound into the fluid. Both the angle of the radiated power as well as the amount of leakage in to the liquid can be varied over quite a wide range taking special measures for the acoustic impedance matching at the respective boundaries. For instance, additional microstructure like gratings or impedance matching layers can be imployed for this purpose. Equally simple and versatile is the generation of a periodic array of such ‚virtual sources', being commensurate with the periodocity of, e.g., a micro titer plate, a microarray or similar small compartements for the sample fluid located on top (or below) of this mixer card.

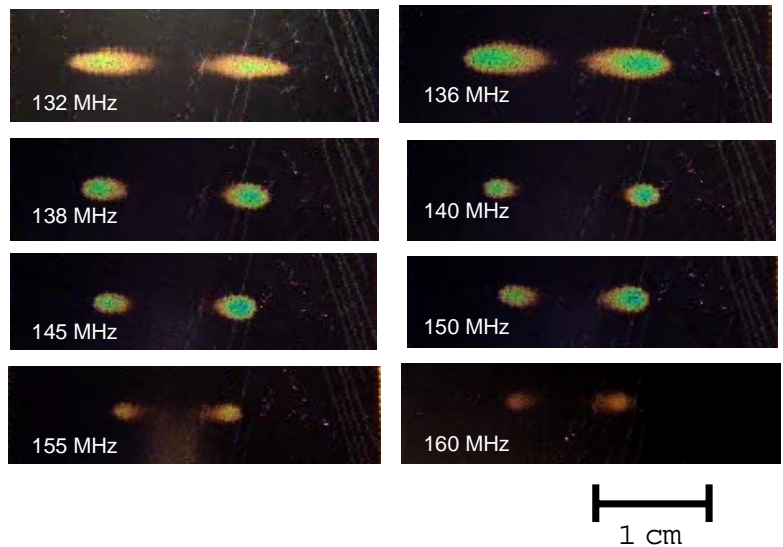

1 cm

**Fig. 4 :** *Visualization of internal ultrasound reflections in a parallel plate ultrasound waveguiding substrate as described in the text. By variation of the frequency of the sound wave, the relative position of the virtual sources can be varied at will. In the figure, only two such sources are shown for clarity.*

In Fig. 4, we depict the result of an experiment where, by variation of the frequency of the ultrasound, the relative position of two of such hot spots has been varied simultaneously. The initial ultrasonic wave was generated using an interdigital transducer on a $LiNbO_3$ substrate being glued to the guiding substrate. This guiding substrate in this case was a 2.85 mm thick non-hardened float glas slide. To minimize internal losses, various materials have been tested, including single crystal quarz, quarz glass, various industrial glasses etc. As expected, the less internal losses occur, the more virtual sound emitters, i.e. internal reflections can be generated. The losses to the fluid, which are intended to generate acoustic streaming inside the liquid layer, also contribute to the achievable pattern. Here, too, the impedance difference between the fluid and the guiding substrate, are responsible for the actual energy transfer per virtual source.

In Fig. 5 and Fig. 6, finally, we show the effect of acoustic streaming to small amounts of fluid contained in the wells of a 384 well micro titer plate (Fig. 5) and a multiwell slide (Fig. 6). To visualize the streaming and the subsequent mixing, small amounts of ink have been added to the buffer solution. The ultrasound in the case of the MTP is coupled to the wells from the bottom of the well, using special impedance matching layers. In the figure, however, only a single well has been adressed ultrasonically. The snapshots are taken approximately ten seconds apart from eachother. As can be seen, the acoustic streaming in the well leads to an agitation and a subsequent mixing of the buffer/ink solution after a couple of seconds. Using the technique of distributed sources by internal reflection of ultrasound, a parallel processing of many wells can be achieved.

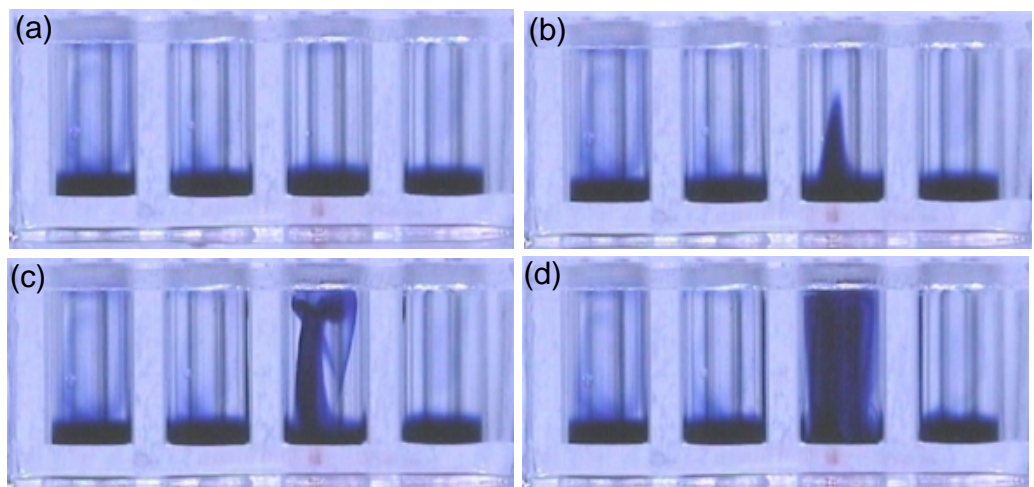

**Fig. 5 :** *Time series of snapshots of the ultrasonically induced mixing in a well of a 384 well micro titer plate. The images are taken approximately ten seconds apart from each other. Ultrasonic energy is selectively coupled into the liquid in a single well from the buttom of the well. Acoustic streaming leads to a rapid and controlled mixing of the liquid as visualized by a small amount of ink being dissolved during agitation.*

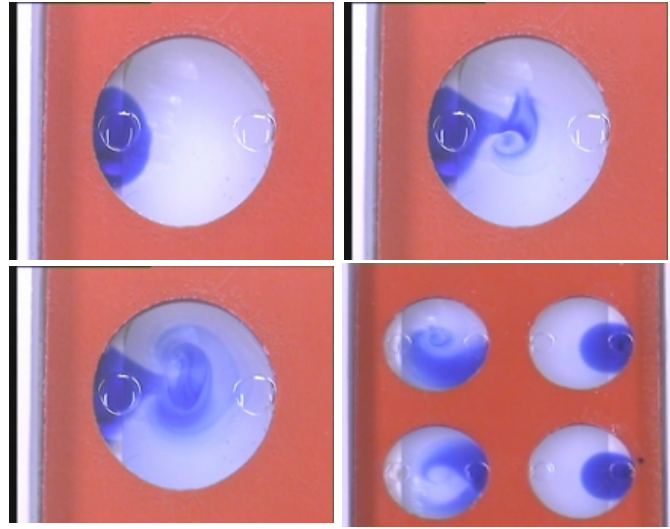

**Fig. 6 :** *Acoustically driven agitation of a small liquid volume in a multiwell slide. In this case, the acoustic energy is coupled to the liquid sideways, using surface acoustic waves as described in the text. The lower right picture shows a section of such a multiwell slide containing 4 separate wells, two of which are being agitated.*

In Fig. 6, we show mixing of a buffer/ ink mixture in a multiwell slide. Here, the reservoirs are defined by openings in a spacer foil, which has been glued on top of a conventional microscope slide. A cover lid having two openings for pipetting the sample defines the volume of the microwell. The ultrasound in this case is coupled in sideways, as can be seen from the evolving streaming pattern.

In summary, we describe a technique employing the effect of acoustic streaming for the agitation of small liquid volumes in thin capillary layers, in multiwell slides, microtiter plates and similar containers forhybridization assays. It turns out to be favourable to distribute several sources of ultrasonic energy over the spotted area of a microarray, or in the case of multiple small containers like micro titer plates, or multi well slides. We demonstrate that it is possible to distribute the sources by taking advantage of internal reflections in an ultrasonic waveguiding layer, hence generating distributed virtual sources in a predetermined manner. The energy losses of a sound wave in a solid to the liquid can be taylored at will, including the adjustment of the geometrical distribution and number of the virtual sources, as well as the amount of energy transferred.
We believe that this technique to mix smalles amounts of liquid by employing our ultrasonic techniques is advantageous over many other approaches and will result in a new generation of instrumentation for biological or chemical assays.

We gratefully acknowledge useful discussions with Christoph Gauer and Jürgen Scriba. Financial support of this investigation has been provided by the Advalytix AG, Brunnthal, Germany.